\documentstyle[]{mn}
\input{psfig}

\newif\ifAMStwofonts

\def\sr{\scriptscriptstyle\rm}

\ifoldfss
  \newcommand{\rmn}[1] {{\rm #1}}

  \ifCUPmtlplainloaded \else
    \NewTextAlphabet{textbfit} {cmbxti10} {}
    \NewTextAlphabet{textbfss} {cmssbx10} {}
    \NewMathAlphabet{mathbfit} {cmbxti10} {} 
    \NewMathAlphabet{mathbfss} {cmssbx10} {} 
  \fi
  \ifAMStwofonts
    \ifCUPmtlplainloaded \else
      \NewSymbolFont{upmath} {eurm10}
      \NewSymbolFont{AMSa} {msam10}
      \NewMathSymbol{\upi}     {0}{upmath}{19}
      \NewMathSymbol{\umu}     {0}{upmath}{16}
      \NewMathSymbol{\upartial}{0}{upmath}{40}
      \NewMathSymbol{\leqslant}{3}{AMSa}{36}
      \NewMathSymbol{\geqslant}{3}{AMSa}{3E}

    \fi
  \fi
\fi 

\ifnfssone
  \newmathalphabet{\mathit}
  \addtoversion{normal}{\mathit}{cmr}{m}{it}
  \addtoversion{bold}{\mathit}{cmr}{bx}{it}
  \newcommand{\rmn}[1] {\mathrm{#1}}

  \newmathalphabet{\mathbfit} 
  \addtoversion{normal}{\mathbfit}{cmr}{bx}{it}
  \addtoversion{bold}{\mathbfit}{cmr}{bx}{it}
  \newmathalphabet{\mathbfss} 
  \addtoversion{normal}{\mathbfss}{cmss}{bx}{n}
  \addtoversion{bold}{\mathbfss}{cmss}{bx}{n}
  \ifAMStwofonts
    \ifCUPmtlplainloaded \else
      \UseAMStwoboldmath
      \makeatletter
      \new@mathgroup\upmath@group
      \define@mathgroup\mv@normal\upmath@group{eur}{m}{n}
      \define@mathgroup\mv@bold\upmath@group{eur}{b}{n}
      \edef\UPM{\hexnumber\upmath@group}
      \new@mathgroup\amsa@group
      \define@mathgroup\mv@normal\amsa@group{msa}{m}{n}
      \define@mathgroup\mv@bold\amsa@group{msa}{m}{n}
      \edef\AMSa{\hexnumber\amsa@group}
      \makeatother
      \mathchardef\upi='0\UPM19
      \mathchardef\umu='0\UPM16
      \mathchardef\upartial='0\UPM40
      \mathchardef\leqslant='3\AMSa36
      \mathchardef\geqslant='3\AMSa3E
    \fi
  \fi
\fi 

\ifnfsstwo
  \newcommand{\rmn}[1] {\mathrm{#1}}

  \DeclareMathAlphabet{\mathbfit}{OT1}{cmr}{bx}{it}
  \SetMathAlphabet\mathbfit{bold}{OT1}{cmr}{bx}{it}
  \DeclareMathAlphabet{\mathbfss}{OT1}{cmss}{bx}{n}
  \SetMathAlphabet\mathbfss{bold}{OT1}{cmss}{bx}{n}
  \ifAMStwofonts
    \ifCUPmtlplainloaded \else
      \DeclareSymbolFont{UPM}{U}{eur}{m}{n}
      \SetSymbolFont{UPM}{bold}{U}{eur}{b}{n}
      \DeclareSymbolFont{AMSa}{U}{msa}{m}{n}
      \DeclareMathSymbol{\upi}{0}{UPM}{'19}
      \DeclareMathSymbol{\umu}{0}{UPM}{'16}
      \DeclareMathSymbol{\upartial}{0}{UPM}{'40}
      \DeclareMathSymbol{\leqslant}{3}{AMSa}{'36}
      \DeclareMathSymbol{\geqslant}{3}{AMSa}{'3E}
    \fi
  \fi
\fi 

\ifCUPmtlplainloaded \else
  \ifAMStwofonts \else 
    \def\upi{\pi}
    \def\umu{\mu}
    \def\upartial{\partial}
  \fi
\fi

\def\LaTeX{L\kern-.36em\raise.3ex\hbox{a}\kern-.15em
    T\kern-.1667em\lower.7ex\hbox{E}\kern-.125emX}

\title{The Ages of Pre-main-sequence Stars}

\author[C. A. Tout, M. Livio and I. A. Bonnell]
  {Christopher A. Tout$^{1,2}$\thanks{E-mail: cat@ast.cam.ac.uk}, 
  Mario Livio$^2$ and Ian A. Bonnell$^1$\\
  $^1$Institute of Astronomy, The Observatories, Madingley Road, Cambridge CB3 0HA\\
  $^2$Space Telescope Science Institute, 3700 San Martin Drive,
Baltimore, MD 21218, U. S. A.}

\date{Submitted 1998 November 16.}

\pagerange{\pageref{firstpage}--\pageref{lastpage}}
\pubyear{1996}

\begin{document}

\label{firstpage}

\maketitle

\begin{abstract}
The position of pre-main-sequence or protostars in the
Hertzsprung--Russell diagram is often used to determine their mass and
age by comparison with pre-main-sequence evolution tracks.  On the
assumption that the stellar models are accurate, we
demonstrate that, if the metallicity is known, the mass obtained is a
good estimate.  However, the age determination can be very misleading
because it is significantly (generally different by a factor of two to
five) dependent on the accretion rate and, for
ages less than about $10^6\,$yr, the
initial state of the star.  We present a number of accreting
protostellar tracks that can be used to determine age if the initial
conditions can be determined and the underlying accretion rate has been constant
in the past.  Because of the balance established
between the Kelvin-Helmholtz, contraction timescale and the accretion
timescale a pre-main-sequence star remembers its accretion history.
Knowledge of the current accretion rate, together with an H--R-diagram
position gives information about the rate of accretion in the past but
does not necessarily improve any age estimate.
We do not claim that ages obtained by comparison with these particular
accreting tracks are likely to be any more reliable than those from
comparisons with non-accreting tracks.  Instead we stress the
unreliability of any such comparisons and use the disparities between
various tracks to estimate the likely errors in age and mass
estimates.  We also show how a set of coeval accreting objects do not
appear coeval when compared with non-accreting tracks.  Instead
accreting pre-main-sequence stars of around a solar mass are likely to
appear older than those of either smaller or larger mass.
\end{abstract}


\section{Introduction}
\label{sec1}

The placement of an observed pre-main-sequence star in the theoretical
Hertzsprung--Russell diagram ($\log L$ against $\log T_{\rmn eff}$) is
notoriously difficult because of its sensitivity to distance and
reddening (see for example Gullbring et al. 1998).  In addition
any contribution to the light from the accretion disc itself must be
subtracted and obscuration by circumstellar material accounted for
(see for example Hillenbrand 1997).
Here, by examining
where theory predicts a particular object ought to lie at a given age,
we investigate what properties of a pre-main-sequence star
can be determined if these difficulties can be overcome.
\par
The process by which stars form from their constituent interstellar
material is as relevant to all branches of astrophysics, from planets
to cosmology, as their subsequent evolution.  However our
understanding and, without doubt, our predictive power lag well
behind.  This is partly because stars which are in the process of
formation are more difficult to
observe.  The relative rapidity of the star formation process means
that there are no nearby pre-main-sequence stars
and the fact that they form in denser regions of the interstellar
medium favours observations at wavelengths longer than optical.  It is
only recently that such objects have begun to be observed in
statistically significant numbers (Cohen \& Kuhi 1979).  From a
theoretical point of view, difficulties arise
because much of the process is dynamical and so does not lend itself
well to the one-dimensional models normally employed in stellar
evolution.  On the other hand we can model the hydrostatic inner regions
using the methods normally employed in
stellar evolution and so, with appropriate boundary conditions,
approximate a forming star.
Indeed this kind of pre-main-sequence theory can be said to have begun
alongside stellar evolution itself with the work of Henyey, Lelevier
\& Levee (1955),
restricted to radiative solutions, and Hayashi (1961), with
convection.  These pioneers were able to describe how a spherical
cloud of gas, already in hydrostatic equilibrium, contracts down to the
main sequence as it releases its own gravitational energy.
\par
The question of how the initial hydrostatic sphere forms is further
complicated by two major effects.  First, dynamical processes
must be important in addition to thermal and nuclear and, second, these
can no longer be expected to be spherically or even
oblate-spheroidally symmetric.  Larson (1969) modelled the spherically
symmetric collapse of a gas cloud that is not yet in hydrostatic
equilibrium.  He showed how the central regions collapse first to form
a hydrostatic core on to which the rest of the cloud accretes.  But
this core cannot behave like Hayashi's pre-main-sequence stars because
its surface is no longer exposed to space and the boundary conditions
are different.
Larson introduced shock conditions at the surface of a near
hydrostatic core.
Building heavily on this,
Stahler, Shu and Taam (1980a,~b, 1981)
were able to follow the evolution of the accreting
core.  Such a core, shrouded in its own accreting envelope, remains
invisible as long as it accretes.  Stahler et al. assumed that
accretion and obscuration cease at the same time when the surrounding
material is somehow blown away.  Their stars then descend the classic
Hayashi tracks until they develop radiative envelopes and move on to
the corresponding Henyey track.  However, if the accreting material
does not obscure the entire stellar surface, we are able to see the
star whilst it is still accreting.  It is this latter situation
that we model in this work.  It is likely to arise because material
accreting from far off will have too much angular momentum to fall
radially on to the central core.  Instead it will form an accretion
disc in a plane perpendicular to the angular momentum axis and
fall on to the core only as viscosity allows a small amount of
material to carry the angular momentum outwards.  If the disc reaches
to the stellar surface then the material will accrete only in an
equatorial band.  If, on the other hand, the central core possesses a
magnetic field strong enough to disrupt the inner parts of the disc,
matter might finally flow in along field lines accreting at relatively
small
magnetic poles or thin accretion curtains.  Similar processes are
known to operate in magnetic cataclysmic variables (Warner 1995 for a
review).  In any of these cases most
of the stellar surface is left exposed and free to radiate like a
normal star.
\par
A comprehensive study of such exposed protostellar cores was
made by Mercer-Smith, Cameron and Epstein (1984).  Because more than
the usual insight is needed to elucidate what they actually did this
work has largely been forgotten.  However it turns out that their
accreting tracks qualitatively differ from ours and so it is important to
identify exactly why this is.  Their calculations begin with a
hydrostatic core of $0.0015\,M_\odot$ of apparently $2\,R_\odot$ based
on the dynamical collapse calculations of Winkler and Newman (1980).
We shall argue in section~3 that, at this mass, the core is still
embedded in a rapidly collapsing cloud and that something like
$0.1\,M_\odot$ and $3\,R_\odot$ gives a more realistic representation
of the central core when
non-spherical accretion begins in earnest.  But we shall also show that the
subsequent evolution is not overly sensitive to this initial state.
More significantly, Mercer-Smith et al. require that at least one
quarter of the accretion luminosity be radiated uniformly over the
whole stellar surface, while we claim that it can all be radiated locally
in a disc boundary layer or localised shocks.  This, coupled with
their extreme accretion rates of typically $10^{-5}\,M_\odot\,\rm
yr^{-1}$, most probably accounts for the huge discrepancy between
their and our tracks, manifested by the fact that their standard model
is at a much higher effective temperature for a given mass than ours.
As a consequence, our models evolve smoothly even if the accretion rate
is abruptly changed while theirs relax to a normal Hayashi track
rapidly over about $100\,$yr when accretion is halted.
\par
A careful analysis of the effects of accretion on stellar structure
has been made by Siess \& Forestini (1996) varying a number of the
physical properties of the accreted material relative to the stellar
surface from angular momentum content to internal energy and find that
reasonable values of these parameters have little affect on the
stellar structure.  Siess, Forestini \& Bertout (1997) then went on to
use their formalism to follow a small number of evolutionary
sequences.  They confirmed the lack of sensitivity to their various
parameters except for the dependence on the fraction, $\alpha$, of the
accretion boundary-layer energy released below the stellar
photosphere.  Large values of this parameter are similar to
Mercer-Smith et al.'s formalism while our models correspond to
$\alpha = 0$.  Siess et al.'s models with $\alpha = 0.01$ are indeed
very similar to our tracks when the accretion history is comparable.
\par
We present several evolution tracks for pre-main-sequence stars accreting from
various initial conditions to quantify the accuracy to which age can
be determined.  Most of the tracks are for solar metallicity of $Z =
0.02$.  However measurements of metallicity in Orion's star forming
regions, although highly uncertain, indicate that $Z = 0.001$ may be
more appropriate (Rubin et al. 1997).  Such low metallicity is also
typical of star forming regions in the Large Magellanic Cloud.
We therefore discuss a set of
low-metallicity tracks which demonstrate how both mass and age
determinations from colour--magnitude diagrams depend
critically on a knowledge of metallicity.
Because the stellar mass function dictates that the bulk of stars have
final masses on the low side we restrict our presentation here to
accreting objects of less than $2\,M_\odot$.  We can expect almost all
stars in the star-forming regions with which we may wish to compare
properties to lie below this mass.  Also, as stressed by Palla and
Stahler (1993), the contraction timescales for massive stars are short
compared with accretion timescales so that the accreting tracks will
tend to follow the zero-age main-sequence and the effective pre-main-sequence
life of massive stars is dominated by their early, low-mass evolution.
Indeed at higher masses the accretion timescale becomes long compared
with the nuclear timescale and it is difficult to separate pre-
and post-main-sequence evolution for some stars.
\par
We find that
masses can be fairly well established if the metallicity is known but
that ages are very dependent on the accretion history and the initial
state of the star particularly below $5\times 10^6\,$yr.
However before we can begin to discuss age
determination we must first establish to what this age is relative.

\begin{figure*}
\caption[]{A Hertzsprung--Russell diagram showing constant-mass
pre-main-sequence tracks of solar metallicity ($Z = 0.02$) stars in
the range $0.1$ to~$2.0\,M_\odot$.  Isochrones of ages ranging from
$10^{3}$ to $10^{8}\,$yr are drawn across the tracks.  The models
were begun at radii large enough that these isochrones are not affected by small
displacements of this starting point.  The zero-age main and
deuterium-burning sequences appear as dots logarithmically spaced in mass.}
\label{fig1}
\end{figure*}

\section{The Zero Age}
\label{sec2}
It is very often unclear how to define the zero-age point for a
forming star and of course it is rather uninformative to quote an age $t$
without explaining exactly what we mean by $t = 0$.  For evolved stars
the zero-age main-sequence (ZAMS) provides a convenient starting point from
which we can both begin the evolution and measure the age of the
star.  The ZAMS must then be defined.  Historically a star was started
in a state of hydrostatic and thermal equilibrium with a uniform
initial composition.  In reality a star never actually passes through
this zero-age state because some nuclear burning takes place while a
newly formed star is still contracting to the main sequence.  In
practice, because the thermal-evolution timescale of
pre-main-sequence stars is several orders of magnitude shorter than
the post-main-sequence nuclear timescale, very little of
the initial hydrogen is burnt and a uniform hydrogen abundance
throughout the star is a reasonable approximation.  This is not so for
the catalytic elements of the CNO cycle in sufficiently massive stars
because these elements are driven towards equilibrium during
pre-main-sequence evolution.  Even so, it is possible to define a
zero-age main sequence (see for example Tout et al. 1996) that roughly corresponds
to the minimum luminosity attained as a star evolves from a pre- to
post-main-sequence phase.  Nor is the assumption of uniform abundance
true for the elements involved in
the pp chain, notably deuterium and He$^3$.  However, on the zero-age
main sequence, the pp chain is complete in transforming hydrogen to
He$^4$ at the stellar centres and so the abundances of D and He$^3$
are in equilibrium for given temperatures and number densities and subsequently
need not be followed explicitly.  On the other hand deuterium burning is a
major source of energy in pre-main-sequence stars and is important throughout this work.
\par
Because we can define a ZAMS reasonably uniquely a
good way to measure pre-main-sequence ages would be backwards from
the ZAMS.  However this is not acceptable if one wishes to
measure the time elapsed since the birth of a star, where relatively small
changes in age lead to large excursions in the H--R diagram.  A
similar problem is encountered with
the upper parts of the red giant, and
particularly asymptotic giant, branch but in these cases we regard
absolute age as relatively useless preferring such
quantities as degenerate core mass as a measure of the evolutionary state
(Tout et al. 1997).
\par
The concept of a stellar birthline in the H--R diagram was introduced
by Stahler (1983) as the locus of points at which stars forming from
a spherically accreting cloud would first become visible.
In the model of Stahler, Shu and Taam (1980a,b, 1981) this occurs when
deuterium ignites in the protostellar core and some ensuing wind blows
away the remainder of the accreting cloud which has, up to this point,
shrouded the star itself from view.  With such a theory, a perfect
place to fix the zero age of a pre-main-sequence star would be the
onset of deuterium burning.  Deuterium burning provides pressure
support for the star for a time comparable with the
Kelvin--Helmholtz timescale $\tau_{\sr KH}$, on which it contracts once
deuterium is exhausted.
This timescale is similar to the entire time taken to contract to this
point from any initial state
so that by the time a star begins to contract again,
below the deuterium burning sequence, it is already relatively old and
has a reasonably well defined age.  Here we concern ourselves with
accretion through a disc.  In this case most of the stellar photosphere is
exposed while accretion is still taking place.  Nor do we assume that
accretion ceases at the onset of deuterium burning.  Under such
circumstances there is no reason why stars should not appear above
Stahler's birthline and it is no longer possible to define a birthline
as a locus of maximum luminosity at which pre-main-sequence stars
appear.  However, the interruption of contraction when deuterium
ignites means that we are much more likely to see stars on and below
the deuterium burning sequence than above it.  By definition Stahler's
birthline is more or less coincident with the deuterium burning
sequence and this explains the consistency of observations with the
idea of a birthline.  Unfortunately this apparent birthline is not the
place where stars are born and so an age measured from a zero-age
deuterium burning sequence is too young by an unknown amount
which is normally at least as much as the deuterium burning lifetime.
\par
D'Antona and Mazzitelli (1994) take another approach which is to begin
evolution at a point in the H--R diagram of sufficiently high
luminosity, or equivalently at
sufficiently large radius on a Hayashi track, that $\tau_{\sr KH}$ is
much less than some acceptable error in the age at any later time.
This error might be chosen to be about $100\,$yr.  Such a definition
leads to a well-defined age at any point on a track corresponding to
a constant mass.
For comparison, figure~\ref{fig1} shows such a set of
pre-main-sequence tracks for $M = 0.1$, $0.2$, $0.5$, $1$
and~$2\,M_\odot$ and isochrones fitted to $50$ models in this range.
We describe our models in detail in the following sections but note
that, because we use very similar physics, they do not differ greatly
from those of D'Antona and Mazzitelli.
\par
However stars do continue to accrete long after
their photospheres are exposed and they can be placed in an H--R
diagram.  A star of about $1\,M_\odot$ is most unlikely to have
reached this mass while $\tau_{\sr KH}$ was still small or indeed even
before deuterium exhaustion.  For this reason we take the zero-age
point of each of our tracks to be a point at which the protostellar
core has the mass and radius of a typical self-gravitating fragment of
a protostellar cloud and model the subsequent evolution with ongoing
accretion.  We then investigate how changing these initial
conditions alters the subsequent isochrones in the H--R diagram to get
an idea of how well we can constrain the age of an observed pre-main-sequence star
relative to its birth as a self-gravitating accreting body.

\section{The stellar models}

We construct our stellar models using the most recent version of the
Eggleton evolution program (Eggleton 1971, 1972, 1973).  The equation of state,
which includes molecular hydrogen, pressure ionization and coulomb
interactions, is discussed by Pols et al. (1995).
The initial composition is taken to be uniform with a hydrogen
abundance $X = 0.7$, helium $Y = 0.28$, deuterium $X_{\rm D} =
3.5\times 10^{-5}$ and metals $Z = 0.02$ with the meteoritic mixture
determined by Anders and Grevesse (1989).  Hydrogen burning is allowed
by the pp chain and the CNO cycles.  Deuterium burning is explicitly
included at temperatures too low for the pp chain.  Once the pp chain
is active hydrogen is assumed to burn to He$^4$ via deuterium and
He$^3$ in equilibrium.  The burning of He$^3$ is not explicitly followed.
Opacity tables are
those calculated by Iglesias, Rogers and Wilson (1992) and Alexander
and Ferguson (1994).  An Eddington approximation (Woolley and Stibbs
1953) is used for the surface boundary conditions at an optical depth
of $\tau = 2/3$.  This means that low-temperature atmospheres, in which
convection extends out as far as $\tau \approx 0.01$ (Baraffe et
al. 1995), are not modelled perfectly.  However the effect of this
approximation on
observable quantities is not significant
in this work (see for example Kroupa and Tout 1997).
\par
We assume that material is accreted from a disc on to a thin
equatorial region of the star so that normal photospheric
boundary conditions are appropriate over most of its surface.  This
would also be true even if the inner edge of the disc is magnetically
disrupted and the material funnelled to a few spots or narrow
accretion curtains whose areas represent a relatively small fraction
of the stellar surface.  Because
our models are one-dimensional we must apply these same boundary
conditions over the whole surface.  Similarly we must assume
that accreted material is rapidly mixed over this same complete surface
so that, on accretion of mass $\delta M$, we can add a spherical shell
of mass $\delta M$ with composition equal to the initial, or ambient,
composition.  We note that the photospheric boundary conditions
effectively fix the thermodynamic state of the accreted material to
those conditions over the radiating stellar surface.  This is
equivalent to the assumption that boundary layer shocks or, in the
case of magnetically funnelled accretion, shocks at or just above the
stellar surface remove any
excess entropy from the accreting material and so is not unduly
restrictive.  Ideally we would like to treat this problem in
two-dimensions.  We could then apply different boundary conditions over
the equatorial band or polar spots where accretion is actually taking place.
With current computational power and techniques such models may not be
too far off (Tout, Cannon and Pichon, private communication).

\section{Initial conditions}

We wish to take as an initial model a typical protostellar core of
mass $M_0$ that is self gravitating within a cloud and that has
reached hydrostatic but not yet thermal equilibrium out to a radius
$R_0$.  Additional material beyond $R_0$ may be gravitationally bound
to the star but not yet accreted.  We assume that the core is
spherically symmetric out to $R_0$ and that beyond this radius material sinks on
to a disc from which it is accreted in a thin equatorial band or other
relatively small part of the stellar photosphere.
\par
The technique we use to construct the initial model is fairly
standard.  We take a uniform composition zero-age main-sequence model
of mass $M_0$ and add in an artificial energy generation rate
$\epsilon_{\rm c}$ per unit mass uniformly throughout the star.
Initially $\epsilon_{\rm c}$ is negligible but we gradually increase
it so that the star is slowly driven back up its Hayashi track.  In a
sense $\epsilon_{\rm c}$ mimics the thermal luminosity that would be
released if the star were contracting down the Hayashi track.  These
objects, however, are in thermal equilibrium.  We continue to increase
$\epsilon_{\rm c}$ until the radius of the object is considerably more
than $R_0$.  At this point we may add or subtract mass freely, while
maintaining hydrostatic and thermal equilibrium, and so vary $M_0$.
In this way we can reach masses below the hydrogen burning limit that
would not have a zero-age main-sequence state of their own.  We then
switch off the artificial energy generation and allow the star to
contract down its Hayashi track supported by the usual gravitational
energy release.  When $R = R_0$ we have our initial model.
\par
We choose a protostellar core of $M_0 = 0.1\,M_\odot$ and
$R_0 = 3\,R_\odot$ as our standard initial
model.  This choice of the initial mass and radius of the
pre-main-sequence star is necessarily somewhat arbitrary because it
depends on the pre-collapse conditions and the dynamics of the
collapse process.  We choose a mass and radius taken to represent a
young star at the end of the collapse and spherical infall phase of
evolution at a time when it first becomes optically visible. This will
include the initial protostellar core that forms from the collapse
phase plus any mass that is accreted on to this core before the infall
becomes significantly aspherical.  This happens when the infalling
material has enough angular momentum to force it to collapse towards a
disc
rather than be accreted directly by the protostellar core.  Any
further accretion from this point will be through this circumstellar
disc.
\par
The gravitational collapse of a molecular cloud forms a first
protostellar core when the density becomes large enough to trap the
escaping IR radiation (e.g. Larson~1969). This sets a minimum mass for
opacity limited fragmentation at about $0.01\,M_\odot$ (Low \&
Lynden-Bell 1976, Rees 1976). This minimum mass will be increased by
the material from further out with low angular momentum (along the
rotation axis) plus the matter that has its angular momentum
removed/redistributed by gravitational torques (e.g. Larson 1984) in
the disc on timescales short compared with the free-fall time.  The
accretion of low-angular momentum matter probably increases the
protostar's mass by a factor of 3 (for initially uniform density
collapse).  The accreted disc material can be estimated as that with
dynamical times significantly less than the original free-fall
time. Thus, material within $20-50\,$au should be accreted within $10^3$
years, which corresponds to disc sizes several times larger than the
first core. For initially solid body rotation and uniform cloud
density this translates to a mass at least three times larger.  We
thus estimate our initial mass as about $0.1\,M_\odot$. This
compares well to the mass of a protostellar core from a spherical
collapse within a fraction of a free-fall time (Winkler \&
Newman 1980) and to the mass within $50\,$au in a collapse including
rotation (Boss~1987, see also Lin \& Pringle~1990).
An initial mass of $0.1\,M_\odot$
is also comparable to the observed lower limit for stellar masses and
still allows for significant mass increase through subsequent accretion.
\par
The choice of the initial stellar radius is perhaps more constrained.
Estimates of this radius depend on the dynamics of the collapse, but
are generally in the range of $2.5$ to $3\,R_\odot$ (Stahler~1988, Winkler \&
Newman~1980). We have chosen the value of $3\,R_\odot$ as given by
Winkler \& Newman (1980) for
an accreting protostar of $0.1$ to greater than $0.5\,M_\odot$.
We investigate variations in $R_0$
and $M_0$ and find that the precise choice is not terribly critical anyway.

\begin{figure*}
\caption[]{A Hertzsprung--Russell diagram showing, as thin lines, tracks followed by
pre-main-sequence stars of initial mass $0.1\,M_\odot$ and radius $3\,R_\odot$
evolved with constant accretion rates ranging from
$10^{-9}$ to $10^{-8}$ in steps of $10^{0.5}$ and then to
$10^{-5.5}\,M_\odot\,\rm yr^{-1}$ in steps of $10^{0.25}$ to a final
mass of $2\,M_\odot$.  Thick lines are
isochrones of $10^4$ to $10^8\,$yr or join points of equal
mass from $0.1$ to $2.0\,M_\odot$.  The zero-age main and
deuterium-burning sequences appear as dots logarithmically spaced in mass.}
\label{fig2}
\end{figure*}

\begin{figure*}
\caption[]{The isochrones and equal-mass loci from figure~\ref{fig2}
-- solid lines -- overlaid with the pre-main-sequence tracks and
isochrones of figure~\ref{fig1} -- dotted lines.  The zero-age main
and deuterium-burning
sequences appear as dots logarithmically spaced in mass.}
\label{fig3}
\end{figure*}

\section{Standard models}

From the standard initial conditions $M_0 = 0.1\,M_\odot$ and $R_0 =
3\,R_\odot$ we evolve a set of thirteen pre-main-sequence stars accreting at
constant rates of $10^{-5.5}$, $10^{-5.75}$, $10^{-6}$, $10^{-6.25}$,
$10^{-6.5}$, $10^{-6.75}$, $10^{-7}$, $10^{-7.25}$, $10^{-7.5}$, $10^{-7.75}$,
$10^{-8}$, $10^{-8.5}$ and $10^{-9}\,M_\odot{\rm yr}^{-1}$.
This range of accretion spans that necessary to produce stars of
between $0.1$ and $3\,M_\odot$ within $10^6\,$yr.  The lower
rates are similar to those observed in classical T~Tauri stars
(e.g. Gullbring et. al. 1998), while the higher rates correspond to
a time average when episodic FU~Ori-type events are
responsible for the bulk of the mass accretion. These outbursts, with
accretion rates of about $10^{-4}\,M_\odot{\rm yr}^{-1}$ lasting about
$100\,$yr and probably recurring every $1{,}000\,$yr
(Hartmann 1991, Kenyon 1999), are necessary to reconcile the
very slow accretion, observed in classical T~Tauri stars (Gullbring
et al. 1998) and in the
younger Class~I objects (Muzerolle, Hartmann \& Calvet 1998), with the
higher envelope infall rate (Kenyon et al. 1990). FU~Ori events have been successfully
modelled as outbursts of high disc-accretion (Hartmann 1991, Bell
et al. 1995, Kenyon 1999).  The timescales on which the accretion rate changes
are small as far as the stellar evolution of the underlying
pre-main-sequence star is concerned, much shorter than its Kelvin-Helmholtz
timescale and so are not likely to alter the progress of stellar
evolution.  On the other hand the thermal timescale of the outer
layers is much shorter and so a fluctuating accretion rate does have
an affect that we should investigate but one which we cannot follow
with our existing machinery nor one that is likely to alter the
luminosity and radius of the star as a whole.
\par
These tracks
are plotted in figure~\ref{fig2} as thin lines where, for clarity, we
terminate them at a mass of
$2\,M_\odot$.  The zero-age main
sequence and zero-age deuterium burning sequence are added to this
figure as dots logarithmically placed in mass.  The evolution of a
pre-main-sequence star is governed by three competing processes and their
associated timescales.  These are gravothermal contraction, accretion
and nuclear burning.  In the early stages, when the thermal timescale
$\tau_{\rm KH}$ is much shorter than the accretion timescale $\tau_{\dot
M}$, the star will evolve down a Hayashi track.  If $\tau_{\dot M}
\ll \tau_{\rm KH}$ it will tend to evolve across to increasing
effective temperature to reach a Hayashi track of correspondingly
larger mass.  Nuclear burning interrupts the contraction once the
central temperature is sufficient to ignite a particular reaction and
for as long as
there is sufficient fuel that the nuclear-burning timescale
$\tau_{\rm N} > \tau_{\rm KH}$.  This
first occurs when deuterium ignites and each track is seen to evolve
up the deuterium-burning sequence by an amount that depends on how
much mass can be accreted before the deuterium is exhausted.  Because
these stars are all fully convective, all the deuterium in the
star is available for burning, as is any additional deuterium that is
accreted at the surface.  Subsequently the tracks are again determined
by the competition between $\tau_{\dot M}$ and $\tau_{\rm KH}$ but
while $\tau_{\dot M}$ is only increasing linearly $\tau_{\rm KH}$ is
growing almost exponentially with time so that accretion eventually
dominates, driving the stars to higher temperature
along tracks parallel to the main sequence.  At the lowest accretion
rates stars descend Hayashi tracks to the zero-age main sequence and
then evolve along it because $\tau_{\dot M} \ll \tau_{\rm N}$.  Above
$0.3\,M_\odot$ stars develop a radiative core on and just above the
main sequence.  This causes a contracting pre-main-sequence star to leave its Hayashi
track and move to higher temperature before reaching the zero-age main
sequence.  At the highest accretion rates $\tau_{\dot M}
\ll \tau_{\rm KH}$ and the stars remain above the main
sequence until $M > 2\,M_\odot$.  Each of these features is
qualitatively described by Hartmann, Cassen \& Kenyon (1997) and a
direct comparison with their work can be found in Kenyon et al. (1998).
\par
Isochrones, or loci of equal age measured from the zero-age point at
$M = M_0$ and $R = R_0$, are drawn as thick lines across the tracks at
ages of $10^4$, $10^5$, $10^6$, $2\times 10^6$, $5\times 10^6$, $10^7$
and $10^8\,$yr.  The $10^4\,$yr isochrone lies so close to the initial
point, and is consequently so prone to small changes in $M_0$ or $R_0$,
that it is futile to claim measurements of age below $10^5\,$yr even
if the initial conditions and accretion rate can be estimated.
The $10^8\,$yr isochrone closely follows the zero-age main sequence.
Loci where the masses are $0.2$, $0.5$, $0.8$, $1.0$
and $2.0\,M_\odot$ are interpolated with thick lines.  The actual
pre-main-sequence track of a $0.1\,M_\odot$ star is drawn as a thick
line to complete the figure.
\par
In figure~\ref{fig3} we overlay both the isochrones and the equal-mass
loci on the pre-main-sequence tracks and isochrones of
figure~\ref{fig1} for comparison.
During the Hayashi contraction the accreting
equal-mass loci follow very closely the tracks of non-accreting
pre-main-sequence stars so that determination of a pre-main-sequence star's
precise position in the Hertzsprung--Russell diagram gives an equally
precise estimate of its mass, irrespective of age and accretion rate.
However, once the radiative core forms, an error in the mass
determination is introduced.
By comparing the change in the slope of the equal-mass loci with the 
non-accreting tracks we see that accretion 
delays the effects of the establishment of the radiative core.
We note that, up to the point at which the radiative core forms,
the fully convective structure has meant that stars evolve essentially
homologously.  For this reason the approximations used by Hartmann et
al. (1997) have remained valid and we expect their tracks to be a good
representation.  Once this homology is lost their equation~(6) is no
longer valid nor is their assumption that luminosity is a unique
function of mass and radius.  The tracks would begin to deviate
radically and it becomes much more important to follow the full
evolution as we do here.
\par
Below $0.2\,M_\odot$ and ages greater
than $10^6\,$yr the isochrones are quite similar too but they deviate
drastically for larger masses with the accreting stars appearing
significantly older than they are by factors of two or more according to
the non-accreting pre-main-sequence isochrones.
This is because, for a given Kelvin-Helmholtz timescale, a lower-mass
Hayashi-track star is smaller in radius so that, as mass is added and
the star moves to higher temperature it always remains smaller and
appears older than a star that originally formed with its current mass.
As an
example, a pre-main-sequence star accreting at $10^{-6.75}\,M_\odot\,\rm yr^{-1}$
would appear to be over $10^7\,$yr old at $1\,M_\odot$ when it is only
$5\times 10^6\,$yr old, while one accreting at $10^{-6}\,M_\odot\,\rm
yr^{-1}$ at $1\,M_\odot$ would appear to be $5\times 10^6\,$yr old when it
is only $2\times 10^6\,$yr old.  Then, because of the delayed
appearance of a radiative core, more massive stars begin to appear
younger again.  For instance, our star accreting at
$10^{-7}\,M_\odot\,\rm yr^{-1}$ would appear to remain at $10^7\,$yr from
about $1.3$ to~$1.74\,M_\odot$, during which time it ages from $1.3$
to~$1.65\times 10^7\,$yr.  At ages of less than $10^6$ years the
position in the Hertzsprung--Russell diagram depends far more on the
starting point than on whether or not the star is accreting but
generally the age will be overestimated by up to the estimate itself.

\begin{figure*}
\caption[]{As Fig.~\ref{fig2} but for
pre-main-sequence stars of initial mass $0.1\,M_\odot$ and radius such that the
initial model is just about to cross
the deuterium burning sequence.}
\label{fig4}
\end{figure*}

\begin{figure*}
\caption[]{The isochrones and equal-mass loci of figure~\ref{fig4} --
solid lines --
overlaid with our standard ones from figure~\ref{fig2} -- dotted lines.}
\label{fig5}
\end{figure*}

\begin{figure*}
\caption[]{As Fig.~\ref{fig2} but for
pre-main-sequence stars of initial mass $0.05\,M_\odot$ and radius $2.4\,R_\odot$.}
\label{fig6}
\end{figure*}

\begin{figure*}
\caption[]{The isochrones and equal-mass loci of figure~\ref{fig6} --
solid lines --
overlaid with our standard ones from figure~\ref{fig2} -- dotted lines.}
\label{fig7}
\end{figure*}

\section{Changing $R_0$ and $M_0$}

To illustrate the sensitivity of the evolutionary tracks, and the
corresponding mass and age determinations, to our choice of initial model
we consider two alternative
starting points.  First, keeping $M_0 = 0.1\,M_\odot$, we reduce $R_0$
until the initial protostellar core is just beginning to ignite
deuterium.  This occurs at about $R_0 = 1.1\,R_\odot$.  Second we
reduce $M_0$ to $0.05\,M_\odot$ and change $R_0$ to $2.4\,R_\odot$ so
as to keep the same mean density.  The evolutionary tracks followed by
pre-main-sequence stars accreting at the same accretion rates, in the range
$10^{-9}$ to~$10^{-5.5}\,M_\odot\,\rm yr^{-1}$, together with
isochrones and equal-mass loci are plotted in figures \ref{fig4}
and~\ref{fig6} respectively.  A non-accreting pre-main-sequence track
for $0.05\,M_\odot$ in figure~\ref{fig6} is seen to run asymptotically
down the main sequence.  Because its mass is below the hydrogen
burning limit of about $0.08\,M_\odot$ this downward progress to lower
luminosities and temperatures will not be halted and this star will
end up as a degenerate brown dwarf.
\par
The isochrones and equal-mass loci are overlaid with those
corresponding to our standard tracks in figures \ref{fig5}
and~\ref{fig7}.  In neither case are the equal-mass loci significantly
affected.  Thus the deviation in these loci from non-accreting
pre-main-sequence tracks can be solely attributed to the
accretion.  On the other hand, it is inevitable that the isochrones are
affected but, in both cases, the difference is small compared with the
overall effect of accretion.  In the case of reduced $R_0$ the
difference is a constant offset of $10^6\,$yr.  This is just the time
taken by a $0.1\,M_\odot$ star to contract from $3$ to
$1.1\,R_\odot$.  Thus ages of about $5\times 10^6\,$yr would be
accurately estimated to within $20$ per cent and those of about
$10^7\,$yr to within $10$ per cent etc.  This error represents the
extreme if we can be sure that all stars are born before igniting
deuterium.  If $R_0$ were increased beyond $3\,R_\odot$ the initial
thermal timescale would be so short that the time taken to contract to
the point of deuterium ignition would not be noticeably different.
\par
From figure~\ref{fig7} we can see that the deviations from our
standard isochrones are even smaller when $M_0 = 0.05\,M_\odot$.  In
this case, what is important is the time taken to accrete the
additional $0.05\,M_\odot$ and so the most slowly accreting tracks are
most affected.  Thus the $10^{-9}\,M_\odot\,\rm yr^{-1}$ track reaches
$0.1\,M_\odot$ after $5\times 10^7\,$yr leading to a relatively large
absolute difference in the $10^8\,$yr isochrone at $0.1\,M_\odot$.
However, as this difference is always relative to the accretion rate, it
rapidly becomes insignificant as we go up in mass.
\par
In both these cases it is important to note that the actual tracks
followed for a given accretion rate become very similar to our
standard ones with the relative time between two points on a track being the same.
It is just the time taken to reach an equivalent point
that alters the isochrones.  We deduce that accretion rate has a much
more significant effect on the position in the H--R diagram than do
the initial conditions.

\begin{figure*}
\caption[]{Three stars accreting at a constant rate of
$10^{-7}\,M_\odot\,\rm yr^{-1}$ from $0.5\,M_\odot$ upwards but
reached by different routes from the same
initial core of $M_0 =
0.1\,M_\odot$ and $R_0 = 3\,R_\odot$: solid line -- constant  $\dot M =
10^{-7}\,M_\odot\,\rm yr^{-1}$; dashed line -- linearly increasing
from $\dot M = 0$ at $t = 0$ to $\dot M = 10^{-7}\,M_\odot\,\rm
yr^{-1}$ at $t = 8\times 10^6\,$yr when $M = 0.5\,M_\odot$;
dotted line -- linearly decreasing rate from $\dot M =
10^{-6}\,M_\odot\,\rm yr^{-1}$ at $t = 0$ to $\dot M =
10^{-7}\,M_\odot\,\rm yr^{-1}$ at $t = 7.3\times 10^5\,$yr when $M = 0.5\,M_\odot$.
Open circles indicate masses of $0.5$ and $1\,M_\odot$
for each of these tracks.  Thin solid lines are our
standard accreting tracks for $10^{-7.5}$, $10^{-6.5}$ and $10^{-6}\,M_\odot\,\rm yr^{-1}$
and non-accreting tracks of $0.5$ and $1\,M_\odot$ while the dashed
line is the locus of $1\,M_\odot$ for the standard accreting tracks.
Dots mark the zero-age
sequences as elsewhere.}
\label{fig8}
\end{figure*}

\section{Variable accretion rates}

The models we have presented so far have accreted at a constant rate
throughout their pre-main-sequence life.  This is unlikely to be the
case in reality.  Even so we might hope that a star of a given mass
and accretion rate might be found at the intersection of the
appropriate accreting track and equal-mass locus of figure~\ref{fig2}
irrespective of its accretion history.  However we find that this is
not the case because a pre-main-sequence star remembers
its past.  
We illustrate the effect of variable accretion rates by
considering three paths, beginning at the same point, that converge to an accretion
rate of $10^{-7}\,M_\odot\,\rm yr^{-1}$ when the star's mass reaches $0.5\,M_\odot$ and continue
to accrete at that rate, constant thereafter.  These tracks are plotted in
figure~\ref{fig8}.  The first has the standard constant accretion rate
of $10^{-7}\,M_\odot\,\rm yr^{-1}$.  The second accretes at a rate
that decreases linearly from $10^{-6}\,M_\odot\,\rm yr^{-1}$ at $t =
0$ to $10^{-7}\,M_\odot\,\rm yr^{-1}$ at $0.5\,M_\odot$ while the third
has an accretion rate that increases linearly from nothing to
$10^{-7}\,M_\odot\,\rm yr^{-1}$.  At $0.5\,M_\odot$ we find that the
stars are well separated in luminosity.  This is to be expected
because a star will take a Kelvin-Helmholtz timescale to adjust its
structure to a new accretion rate.  As discussed in section~5, it is
this same timescale that is balanced with the accretion timescale that
is directly responsible for the deviation of the tracks.  This
timescale balance will be maintained until nuclear burning becomes
important, in this case on the zero-age main sequence.  Consequently
the stars are never given enough time to thermally relax and their
early accretion history can be remembered throughout their
pre-main-sequence evolution.
\par
At $1\,M_\odot$ the mass can still be estimated accurately by
comparison with the standard accreting tracks of figure~\ref{fig2} but
note that this differs from the mass that would be
estimated if we were to compare with non-accreting tracks.  At
$0.5M_\odot$ age estimates from non-accreting tracks would be between
30 and~60 per cent too old and at $1\,M_\odot$ between 2 and~3 times
too old for each of these stars.  This reflects the general nature and
magnitude of the difference between non-accreting and
our standard accreting pre-main-sequence stars, comparison with
which would give a better estimate in each of these particular cases
(within 10 per cent at $0.5\,M_\odot$ and 20 per cent at $1\,M_\odot$).
\par
We emphasize again that we cannot estimate the current accretion
rate from the position in the H--R diagram but if we know this current
rate then placement in an H--R diagram does give us information about
the accretion history.  This behaviour is in accord with equation~(6)
of Hartmann et al. (1997), with $\alpha = 0$.  As long as $\dot M/M$
dominates $\dot R/R$, this equation predicts $R$ as a function of $M$
and $\dot M$ only.  Our track with decreasing $\dot M$ always has $\dot
R/R$ somewhat greater than $\dot M/M$ because it has reached $M =
0.5\,M_\odot$ with the two terms in balance at higher accretion rates.
\begin{figure*}
\caption[]{A Hertzsprung--Russell diagram showing constant-mass
pre-main-sequence tracks of low metallicity ($Z = 0.001$) stars in
the range $0.1$ to~$2.0\,M_\odot$.  Isochrones of ages ranging from
$10^{3}$ to $10^{8}\,$yr are drawn across the tracks.  The models
were begun at radii large enough that these isochrones are not affected by small
displacements of this starting point.  The zero-age main and
deuterium-burning sequences appear as dots logarithmically spaced in mass.}
\label{fig9}
\end{figure*}

\begin{figure*}
\caption[]{The pre-main-sequence tracks and isochrones for $Z=0.001$
from figure~\ref{fig9} -- solid lines -- overlaid with those for solar
metallicity from figure~\ref{fig1} -- dotted lines.}
\label{fig10}
\end{figure*}

\begin{figure*}
\caption[]{A Hertzsprung--Russell diagram showing, as thin lines, tracks followed by
pre-main-sequence stars of initial mass $0.1\,M_\odot$ and radius $3\,R_\odot$ and
metallicity $Z = 0.001$
evolved with constant accretion rates ranging from
$10^{-9}$ to $10^{-5.5}\,M_\odot\,\rm yr^{-1}$.  Thick lines are
isochrones of $10^4$ to $10^7\,$yr or join points of equal
mass from $0.1$ to $2.0\,M_\odot$.  The zero-age main and
deuterium-burning sequences appear as dots logarithmically spaced in mass.}
\label{fig11}
\end{figure*}

\begin{figure*}
\caption[]{The isochrones and equal-mass loci from figure~\ref{fig11}
-- solid lines -- overlaid with the pre-main-sequence tracks and
isochrones of figure~\ref{fig9} -- dotted lines.  The zero-age main
and deuterium-burning
sequences appear as dots logarithmically spaced in mass.}
\label{fig12}
\end{figure*}

\section{Changing metallicity}

Finally we consider the effect of different metallicities.  In general
reducing the metallicity moves the zero-age main sequence to hotter
effective temperatures and slightly higher luminosities (see for
example Tout et al. 1996).  This is due to decreased opacity when
there are fewer metal atoms providing free electrons.  This shift
is reflected throughout the pre-main-sequence evolution.
Figure~\ref{fig9} shows the same tracks and isochrones as
figure~\ref{fig1} but for a metallicity of $Z = 0.001$.  These models
have an initial helium abundance of $Y = 0.242$ and hydrogen $X =
0.757$ to account for the less-processed interstellar medium from
which such stars must be forming.  In practice we should
correspondingly increase the deuterium abundance too but, because this
is very uncertain anyway, we leave it at $X_{\rm D} = 3.5\times
10^{-5}$ so as not to convolute the differences between the two
metallicities.  Apart from the shift, the tracks are qualitatively
similar except for the disappearance of the second hook just above the
ZAMS in the more massive star tracks.  At $Z = 0.02$ this is due
to the CNO catalytic isotopes moving towards equilibrium in the
stellar cores before hydrogen burning begins in earnest.
\par
Figure~\ref{fig10} overlays these tracks and isochrones with those for
$Z = 0.02$.  We can see directly that an error of a factor of two or
more would be made in the mass estimate and a factor of ten or so in the age if a
pre-main-sequence star of metallicity $Z = 0.001$ were compared with
models made for $Z = 0.02$.  Clearly, if stars are indeed still forming at
such low metallicities, it is very important to be sure of the precise
value before making any comparisons.  For a rough estimate of how the
tracks move with metallicity we
interpolate these two sets of tracks together with a
similar set for $Z = 0.01$.  We find the difference in mass
\begin{equation}
\delta M = M(Z=0.02) - M(Z)
\end{equation}
between tracks of metallicity $Z$ and those of solar metallicity that
pass through a given point $(L,T_{\rm eff})$ in the Hertzsprung--Russell
diagram to be
\begin{equation}
\delta M \approx 0.164\left(\log_{10}{Z\over
0.02}\right)^{-0.7}\left(L\over L_\odot\right)^{0.25}\left(T_{\rm
eff}\over 10^{3.6}\,{\rm K}\right)^6
\end{equation}
to within 20 per cent before the evolution turns away from the Hayashi 
tracks.
\par
Figure~\ref{fig11} shows the accretion tracks starting from the
standard initial conditions of $M_0 = 0.1\,M_\odot$ and $R_0 =
3\,R_\odot$, together with the associated isochrones and equal-mass
loci for $Z = 0.001$.  Figure~\ref{fig12} overlays these with the
non-accreting tracks and corresponding isochrones.  Similar comments
can be made concerning the mass and age determinations as for $Z =
0.02$ from figure~\ref{fig3}.  Note however that, at this low
metallicity, all the stars have developed a radiative core before
reaching $2\,M_\odot$.

\section{Conclusions}

\begin{figure*}
\caption[]{The distribution of the factor by which mass at a given
point in the H--R diagram differs between our standard accreting
tracks and non-accreting tracks.  Over most of the diagram the
difference is small being nowhere more than 50 per cent.  At
temperatures less than about $10^{3.76}\,$K the accreting mass is
larger while at higher temperatures the accreting mass is smaller.
Some non-accreting tracks and isochrones are overlayed and the shaded
region is that for which we have both accreting and non-accreting
tracks available.  Twice as many accreting tracks as plotted in
figure~\ref{fig2} were required to achieve the resolution of this
figure.}
\label{fig13}
\end{figure*}

\begin{figure*}
\caption[]{The distribution of the factor by which age at a given
point in the H--R diagram differs between our standard accreting
tracks and non-accreting tracks.  For
temperatures less than about $10^{3.76}\,$K the accreting stars are
younger (i.e. they appear older than they are when their age is
estimated by comparison with non-accreting
tracks) while at higher temperatures they are older.
Some non-accreting tracks and isochrones are overlayed and the shaded
region is that for which we have both accreting and non-accreting
tracks available.  Twice as many accreting tracks as plotted in
figure~\ref{fig2} were required to achieve the resolution of this
figure.}
\label{fig14}
\end{figure*}

If the metallicity of a star forming region is known then the masses
on the Hayashi tracks can be fairly accurately determined.
As noted by Siess et al. (1997)
accretion delays the formation of a radiative core, which consequently
begins further down the Hayashi track at a given mass.
However the locus
of equal mass points will subsequently move to higher luminosities than
a non-accreting star of the same mass.  Thus the mass determined by comparison
with the Henyey portion of any track can be either an under or an
overestimate.
Figure~\ref{fig13} shows the relative error that might be made in
estimating the mass over the region of interest in the H--R diagram.
In the Hayashi region accretion generally leads to an
overestimate of the age in a comparison with non-accreting tracks
while it can lead to an underestimate during the Henyey phase.  At any
time these errors in age could be a factor of two or more.
Figure~\ref{fig14} illustrates the error distribution for age
estimates.
In addition, not knowing the zero-age mass and radius of the star
can lead to an absolute offset in age of up to about $10^6\,$yr so that
any age estimate of less than about $2\times 10^6\,$yr cannot be
trusted.  Ages larger than this can be significantly in error if
accretion is taking place  at an unknown rate but, once accretion has ceased, the age
can be expected to correspond to non-accreting isochrones within a
Kelvin--Helmholtz timescale.  Thus the absolute error would be about equal to the age at
which accretion became insignificant.  
In all other cases great care must be taken when estimating ages.

\begin{figure}
\caption[]{The estimated age and mass for each point along the
$5\times 10^6\,$yr fitted to our standard accreting tracks (Figure~\ref{fig2})
when the luminosity and effective temperature are compared with the
non-accreting tracks of Figure~\ref{fig1}.}
\label{fig15}
\end{figure}

\par
Apart from the initial comparatively small offsets, even if the initial
conditions of a set of stars are known to be the same, relative ages
are equally affected by accretion history.
As
a particular example, we may wish to decide whether two components of
a pre-main-sequence binary star are coeval.
If, as pre-main-sequence
stars often do, one or both lies in the temperature range between
$10^{3.55}$ and $10^{3.75}\,$K where the error in age is likely to be
more than a factor of two we can expect a significant difference in
estimated age even in a coeval system.
This binary example can be extended to star clusters.
Accretion can lead to an apparent mass-dependent age-spread in
otherwise coeval systems when non-accreting pre-main-sequence tracks
are used to estimate ages.  At any time, if all the stars in a cluster
are coeval and began with the same initial core mass, the
low-mass stars must have accreted less and hence have lower
disc-accretion rates than those of higher mass which must have
accreted more material.
Thus accretion does not greatly affect the age
determination of low-mass stars while higher-mass stars are more
affected.  Comparison with non-accreting tracks makes these appear older
while on Hayashi tracks and younger
on the Henyey tracks (see figure~\ref{fig14}).  Thus, intermediate-mass
pre-main-sequence stars can look older than their low-mass
counterparts (by up to a factor of five) while yet higher-mass stars can
appear younger again.  Figure~\ref{fig15} illustrates this point by
plotting the estimated age against the estimated mass for all points
along the $5\times 10^6\,$yr isochrone fitted to our standard accreting
tracks.  Though this is a particular case for a particular set of
models this qualitative behaviour would be true of any coeval sample
that is still undergoing accretion.  Indeed mass-dependent ages have been recorded in
several young stellar clusters where the lowest-mass stars appear
youngest with increasing ages for the intermediate mass stars and
lower ages again for the higher mass stars (Hillenbrand 1997;
Carpenter et. al. 1997).  These mass-dependent ages may reflect
ongoing disc-accretion rather than a dispersion in formation time
and age determinations in these clusters should
be reevaluated in the light of this work.
\par
If the metallicity is not known
the situation becomes even worse.  For instance, as mentioned in the
introduction, the metallicity of some extragalactic star forming
regions, and possibly even Orion, may be
as low as $Z = 0.001$.  This would lead to an overestimate in mass by
a factor of two or more and an overestimate in age by about a factor
of ten if a comparison were inadvertently made with solar metallicity tracks.

\section*{ACKNOWLEDGMENTS}

CAT and IAB are very grateful to PPARC for advanced fellowships.  CAT also thanks
NATO, the SERC and the University of California in Santa Cruz for a
fellowship from 1990-1 when much of the foundation for this work was
laid and the Space
Telescope Science Institute for an eight month position during which
it was continued.  Many thanks go to Jim Pringle for mild but chronic
goading and for many ideas and suggestions along the way.

\label{lastpage}

\end{document}